\documentstyle[12pt]{article}
\def\PRL #1 #2 #3{{\sl Phys. Rev. Lett.} {\bf#1} (#2) #3}
\def\NPB #1 #2 #3{{\sl Nucl. Phys.} {\bf B#1} (#2) #3}
\def\NPBFS #1 #2 #3 #4{{\sl Nucl. Phys.} {\bf B#2} [FS#1] (#3) #4}
\def\CMP #1 #2 #3{{\sl Commun. Math. Phys.} {\bf #1} (#2) #3}
\def\PRD #1 #2 #3{{\sl Phys. Rev.} {\bf D#1} (#2) #3}
\def\PLA #1 #2 #3{{\sl Phys. Lett.} {\bf #1A} (#2) #3}
\def\PLB #1 #2 #3{{\sl Phys. Lett.} {\bf #1B} (#2) #3}
\def\JMP #1 #2 #3{{\sl J. Math. Phys.} {\bf #1} (#2) #3}
\def\PTP #1 #2 #3{{\sl Prog. Theor. Phys.} {\bf #1} (#2) #3}
\def\SPTP #1 #2 #3{{\sl Suppl. Prog. Theor. Phys.} {\bf #1} (#2) #3}
\def\AoP #1 #2 #3{{\sl Ann. of Phys.} {\bf #1} (#2) #3}
\def\PNAS #1 #2 #3{{\sl Proc. Natl. Acad. Sci. USA} {\bf #1} (#2) #3}
\def\RMP #1 #2 #3{{\sl Rev. Mod. Phys.} {\bf #1} (#2) #3}
\def\PR #1 #2 #3{{\sl Phys. Reports} {\bf #1} (#2) #3}
\def\AoM #1 #2 #3{{\sl Ann. of Math.} {\bf #1} (#2) #3}
\def\UMN #1 #2 #3{{\sl Usp. Mat. Nauk} {\bf #1} (#2) #3}
\def\FAP #1 #2 #3{{\sl Funkt. Anal. Prilozheniya} {\bf #1} (#2) #3}
\def\FAaIA #1 #2 #3{{\sl Functional Analysis and Its Application} {\bf
#1} (#2) #3}
\def\BAMS #1 #2 #3{{\sl Bull. Am. Math. Soc.} {\bf #1} (#2)
#3} \def\TAMS #1 #2 #3{{\sl Trans. Am. Math. Soc.} {\bf #1} (#2) #3}
\def\InvM #1 #2 #3{{\sl Invent. Math.} {\bf #1} (#2) #3}
\def\LMP #1 #2 #3{{\sl Letters in Math. Phys.} {\bf #1} (#2) #3}
\def\IJMPA #1 #2 #3{{\sl Int. J. Mod. Phys.} {\bf A#1} (#2) #3}
\def\AdM #1 #2 #3{{\sl Advances in Math.} {\bf #1} (#2) #3}
\def\RMaP #1 #2 #3{{\sl Reports on Math. Phys.} {\bf #1} (#2) #3}
\def\IJM #1 #2 #3{{\sl Ill. J. Math.} {\bf #1} (#2) #3}
\def\APP #1 #2 #3{{\sl Acta Phys. Polon.} {\bf #1} (#2) #3}
\def\TMP #1 #2 #3{{\sl Theor. Mat. Phys.} {\bf #1} (#2) #3}
\def\JPA #1 #2 #3{{\sl J. Physics} {\bf A#1} (#2) #3}
\def\JSM #1 #2 #3{{\sl J. Soviet Math.} {\bf #1} (#2) #3}
\def\MPLA #1 #2 #3{{\sl Mod. Phys. Lett.} {\bf A#1} (#2) #3}
\def\JETP #1 #2 #3{{\sl Sov. Phys. JETP} {\bf #1} (#2) #3}
\def\JETPL #1 #2 #3{{\sl  Sov. Phys. JETP Lett.} {\bf #1} (#2) #3}
\def\PHSA #1 #2 #3{{\sl Physica} {\bf A#1} (#2) #3}
\def\CQG #1 #2 #3{{\sl Class. Quantum Grav.} {\bf #1} (#2) #3}
\def\SJNP #1 #2 #3{{\sl Sov. J. Nucl. Phys. (Yadern.Fiz.)} {\bf #1} (#2) #3}
\def\a{\alpha}\def\b{\beta}\def\g{\gamma}\def\d{\delta}\def\e{\epsilon}

\def\k{\kappa}
\def\Th{\Theta}\def\th{\theta}\def\om{\omega}\def\Om{\Omega}\def\G{\Gamma}
\newcommand{\nn}{\nonumber\\}\newcommand{\p}[1]{(\ref{#1})}
\begin{document}
\thispagestyle{empty}
\renewcommand{\thefootnote}{\fnsymbol{footnote}}
\begin{center}
{\Large\bf
Doubly Supersymmetric Geometric Approach
for Heterotic String: \\
from Generalized Action Principle \\
to Exactly Solvable Nonlinear Equations }\\
\bigskip
\bigskip
{\bf Igor A. Bandos, }
\\
{\sl Kharkov Institute of Physics and Technology, \\
310108, Kharkov, the Ukraine}
\\ e-mail: kfti@rocket.kharkov.ua
\\
\bigskip
\bigskip

{\bf Abstract}
\end{center}

    The previously proposed  generalized action principle approach to
supersymmetric extended objects is considered in some details for the case
of heterotic string in $D=3, 4, 6 ~and~ 10$ space--time dimensions.
    The proof of the 'off--shell' superdiffeomorphism invariance of the
generalized action is presented.
    The doubly supersymmetric geometric
approach to heterotic string is constructed  on the basis of generalized
action principle (instead of the geometrodynamic condition, used for this
previously).

   It is demonstrated that $D=3$ heterotic string is described by $n=(1,0)$
supersymmetric generalization of the nonlinear Liouville equation.

\newpage

\renewcommand{\thefootnote}{\arabic{footnote}}

\begin{center}

{\bf Introduction}

\end{center}

This talk contains some results of the investigations, which were
performed in collaboration  with Dmitrij P. Sorokin and  Dmitrij V. Volkov
and devoted to the development of twistor-- like approach for superstring
and supermembrane theories.

These investigations  are the natural continuation of ones from Ref.
\cite{bsv}, where generalized action principle for supersymmetric extended
objects was proposed, as well as of ones from Ref. \cite{bpstv}, where the
doubly supersymmetric geometric approach for superstrings and
supermembranes was built on the basis of the so--called geometrodynamic
condition (see below).

The main original motivation for the development of
different versions of a twistor--like approach
for superparticles and supersymmetric extended objects
\cite{stv}--\cite{bzp} was the construction of an adequate basis for
future attempts to attack the covariant quantization problem, which
solution  seems to be necessary for deeper understanding of the quantum
theory of supersymmetric extended objects
\footnote{In the standard formulations of the superparticle \cite{bs},
superstrings \cite{gs,gsw} and supermembranes \cite{m} the $\kappa$--
symmetry \cite{k} play a significant role \cite{gsw}, however its meaning
and origin had been unclear \cite{gsw}. Moreover, it has an infinitely
reducible form and its generator can not be splitted covariantly from the
fermionic second class constraints in the Hamiltonian formalism. The later
properties hamper a covariant quantization of superstrings.}.

In a Lorentz--harmonic twistor--like (component) formulation of refs.
\cite{bhbz0}--\cite{bzp}
the $\k$--symmetry was represented in an irreducible but rather
complicated form.

The twistor--like approach based on a superfield
formulation of super--p--branes in world superspace
\cite{stv}--\cite{bers94} allowed one
to replace the $\k$--symmetry by more fundamental local world
supersymmetry and thereby to solve the problem of the infinite
reducibility of the former.

At the same time some basic problems
have not been solved satisfactory in the known versions of the
approach both from the
aesthetic and practical point of view. For instance, for constructing
the superfield action one should use superfield Lagrange multipliers.
Though some
of their components can be identified (on the mass shell) with the
momentum density and the tension of the super--p--brane, in general, the
geometrical and physical meaning of the Lagrange multipliers is obscure.
Moreover, in a version suitable for the description of D=10, 11 objects
\cite{to,dghs92,gs2,tp93,bers94}
 their presence in the action gives rise to some new symmetries
which turn out to be infinite reducible themselves, so that the problem
which we fighted in the conventional Green--Schwarz formulation
reappeared in a new form in the twistor--like formulation. Another point
concerning the Lagrange multipliers is that in the superfield
formulation of D=10 type II superstrings \cite{gs2} and a D=11, N=1
supermembrane \cite{tp93} Lagrange multipliers become propagative redundant
degrees of freedom which may spoil the theory at the quantum level.

All this has forced us to revise the twistor--like superfield approach
on the basis of more geometrically grounded
reasons \cite{bpstv,bsv} and to apply the
generalized action principle of the rheonomic
approach \cite{rheo} to superstrings and supermembrane \cite{bsv}.

The generalized action principle, first of all, gives the possibility to
reproduce the superfield equations of motion for superstrings and other
super--$p$--branes just in the form suitable for the development of the
doubly supersymmetric geometrical approach.

And, hence, it open a new possibility for a natural application of the
doubly supersymmetric twistor-- like approach to the studying of some
(quasi--) classical problems of superstring and supermembrane theories.
Among such problems are ones related to a coupling of the super--$p$--
branes to the natural background (super)fields, including the
investigation of $T$ --duality (see, for example,
 \cite{dual} and refs. therein) in
terms of cotangent bundle \cite{klso}, and investigation of nonlinear
equations of motion of super--$p$-- branes with $p \geq 2$.

In this talk we will consider in details generalized action for $D=3, 4,
6 ~and ~ 10$ heterotic string
 \footnote{For simplicity, we will not consider heterotic
 fermions inputs here, so 'heterotic string' means here a closed
 string with $N=1$ target space supersymmetry.},
 describe its variation and derivation of
the superfield equations of motion. Then we will construct the doubly
supersymmetric geometric approach
\cite{bpstv,bsv} for $D=3, 4, 6 ~and~ 10$ heterotic
string, analyze its equations in the simplest case $D=3$ and prove, that
they can be reduced to $n=(1,0)$ supersymmetric generalization of
nonlinear Liouville equation.

\section{The generalized action principle for
heterotic superstrings}

The basic concepts and properties of generalized action for super--$p$--
branes can be found in Ref. \cite{bsv}. All of them have the counterparts
in the rheonomic approach \cite{rheo} developed for supergravity
(see \cite{vs} -- \cite{v} and refs. in \cite{v}).
However the super-- $p$--brane  case is much more simple since for
constructing the action only the simplest geometrical objects (i.e.
vielbeins, and not connection and curvature) are involved.

So, let us begin from the prescription suitable for the
construction of the generalized action.

-- {\sl step 1} -- we shall find the component superstring action
which is written (or can be rewritten) in terms of
differential forms  without
use of the Hodge operation $*$
\footnote{We will stress, that the standard Green--Schwarz superstring
action \cite{gs,gsw} can not be used for this. Indeed,
even its bosonic limit, which is the Polyakov's string action,
has the form $\int dx^{\underline{m}} * dx^{\underline{m}}$}.

-- {\sl step 2} -- we shall replace in this action all the fields by
superfields and ordinary world--sheet
${\cal M}^2_0$
\begin{equation}\label{m20}
{\cal M}^{2}_0~ = \{ (\xi^{m}, \eta^q); \eta^q = 0 \}
\end{equation}
by an arbitrary two--dimensional bosonic surface ${\cal M}^2$
\begin{equation}\label{m2}
{\cal M}^{2}~ = \{ (\xi^{m}, \eta^q); \eta^q = \eta^q(\xi) \}
\end{equation}
in world--sheet
superspace $\Sigma^{(2|D-2)}$
\begin{equation}\label{wsssp}
\Sigma^{(2|D-2)}~ = \{ (\xi^{m}, \eta^q) \} , \qquad
m= 0, 1  \qquad  q= 1, \ldots, (D-2)
\end{equation}

Fortunately, such component action exists. It is known as the action of the
so--called twistor--like Lorentz harmonic formulation
\cite{bzst}--\cite{bzp}. So, rewriting it as a product of differential
1--forms \cite{bpstv} and then doing with it the {\sl step 2}, we get the
generalized action described below.

\subsection{The action functional}

So, the action for $D=3,~4,~6~and~10$ heterotic string
is
\begin{equation}\label{1}
{\cal S} = \int_{{\cal M}^{2}} ~~~~{\cal L}_2
\end{equation}
where  ${\cal M}^2$ is an arbitrary surface \p{m2} in world sheet
superspace \p{wsssp}, Lagrangian 2--form
\begin{equation}\label{1a}
 {\cal L}_2 = -{1 \over 2}
 \left( E^{++} e^{--} - E^{--} e^{++} + e^{--} e^{++}
\right) - i dX^{\underline{m}}
d\Th\G_{\underline{m}} \Th,
\end{equation}
is constructed out of world sheet superspace bosonic vielbein 1--forms
\begin{equation}\label{2bose}
e^a(\xi,\eta) \equiv (e^{++}(\xi,\eta), e^{--}(\xi,\eta) )
\end{equation}
and two vielbein 1--forms of a flat target superspace
\begin{equation}\label{rep_indb}
E^{\pm\pm} \equiv
\Pi^{\underline{m}}u_{\underline{m}}^{\pm\pm} , \qquad
\end{equation}
by use of exterior product
of the forms without any application of the Hodge operation.

The complete basis of the
superspace cotangent to a  world supersurface (supervielbein) contains
besides \p{2bose} also $(D-2)$ fermionic $1$--forms $e^{+q}(\xi, \eta)$:
\begin{equation}\label{2}
e^A=(e^{++}, e^{--}, e^{+q}) , \qquad  q= 1, \ldots, (D-2)
\end{equation}
but they are not involved into the action explicitly.

Thus $\xi$--directions have a privilege over $\eta$--directions.

However, the external differential $d$ should be expended in the
complete $e^A$ basis
\begin{equation}\label{d}
d=e^a \nabla_a+e^{\a p} \nabla_{\a p} ,
\end{equation}
where $\nabla_{\pm\pm},~\nabla_{+ q}$ are covariant
derivatives for world--sheet scalar superfields.

The forms
\begin{equation}\label{pif}
\Pi^{\underline{m}}=dX^{\underline{m}}-id\Th\G^{\underline{m}}\Th,\qquad
d\Th^{\underline{\mu}} ,
\end{equation}
involved into \p{rep_indb} and \p{1a}, are
the  pullbacks onto the world sheet superspace
$\Sigma^{(2|D-2)}$
of the basic
supercovariant forms \cite{volkov} of flat target superspace.

$u_{\underline{m}}^{++}(\xi,\eta)$,
$u_{\underline{m}}^{--}(\xi,\eta)$, involved in Eq.\p{rep_indb},
are the light--like vector components
of a local frame (supervielbein)
\begin{eqnarray}\label{rep_ind}
E^{\underline A} \equiv
( E^{\underline a}; E^{\underline \a})
\equiv ( E^{++}, E^{--}, E^{i};
E^{\a q}, E^{\a \dot q}) \nn
E^{ \pm\pm} \equiv
\Pi^{\underline{m}}u_{\underline{m}}^{\pm\pm} , \qquad
E^{i} \equiv
\Pi^{\underline{m}}u_{\underline{m}}^{~i} , \qquad \nn
E^{\underline \a} \equiv
d\Theta^{\underline{\mu}} v_{\underline{\mu}}^{~\underline \a} , \qquad
\end{eqnarray}
in target superspace.
Together with the $(D-2)$ components
$u_{\underline{m}}^{i}(\xi,\eta)$ they
are naturally \cite{bhbz0,ghs,bzst,bzm,bzp,bpstv}
composed of the spinor moving frame matrix components
(Lorentz harmonics or generalized Newman-- Penrose dyades)
\begin{equation}\label{harm}
v_{\underline{\mu}}^{~\underline{\a}} =
( v^{~+}_{\underline{\mu} q}~,~v^{~-}_{\underline{\mu}\dot q} )
{}~~~~\in ~~~Spin(1,D-1) \qquad
\end{equation}
as follows
\begin{eqnarray}\label{10D}
u^{++}_{\underline{ m}}
\Gamma^{\underline{ m}}_{\underline{ \mu} \underline{ \nu}}
= 2 v^{~+}_{\underline{\mu} q} v^{~+}_{\underline{\nu} q} , \qquad
u^{--}_{\underline{ m}}
\Gamma^{\underline{ m}}_{\underline{ \mu} \underline{ \nu}}
= 2 v^{~-}_{\underline{\mu} \dot q}
v^{~-}_{\underline{\nu} \dot q} , \qquad \nn
u^{~i}_{\underline{ m}}
\Gamma^{\underline{ m}}_{\underline{\mu} \underline{\nu}}
= ( v^{~+}_{\underline{\mu} q}
v^{~-}_{\underline{\nu} \dot q}
+ v^{~-}_{\underline{\mu} \dot q}
v^{~+}_{\underline{\nu} q}) \g^{i}_{q \dot q}
, \qquad \nn
u^{++}_{\underline{ m}}
\Gamma^{\underline{ m}~\underline{ \mu} \underline{ \nu}}
= 2 v^{+\underline{\mu}}_{\dot q} v^{+\underline{\nu}}_{\dot q} ,
\qquad
u^{--}_{\underline{ m}}
\Gamma^{\underline{ m}~\underline{ \mu} \underline{ \nu}}
= 2 v^{-\underline{\mu}}_{q} v^{-\underline{\nu}}_{q} , \qquad \nn
u^{~i}_{\underline{ m}}
\Gamma^{\underline{ m}~\underline{\mu} \underline{\nu}}
= - ( v^{-\underline{\mu}}_{q} v^{+\underline{\nu}}_{\dot q}
+ v^{+\underline{\mu}}_{\dot q} v^{- \underline{\nu}}_{q})
\g^{i}_{q \dot q} , \qquad
\end{eqnarray}
In \p{10D} we have
presented the considered expressions for the case of target
superspace with $D=10$, where the $SO(1,1) \otimes SO(8)$ invariant
representation for the (chiral) gamma matrices
 $\G_{\underline{a}}$ has the form
\begin{eqnarray}\label{g10D}
\G^{++}_{\underline{\a}\underline{\b}} =
\left(\matrix{2 \d_{qp} & 0\cr 0 & 0\cr} \right)
= \G^{--~\underline{\a}\underline{\b}} , \qquad \nn
\G^{--}_{\underline{\a}\underline{\b}} =
\left(\matrix{0 & 0\cr 0 & 2 \d_{\dot{q}\dot{p}} \cr} \right)
= \G^{++~\underline{\a}\underline{\b}} , \qquad \nn
\G^{i}_{\underline{\a}\underline{\b}} =
\left(\matrix{0 & \gamma ^{i}_{q \dot p}\cr
\tilde{\gamma}^{i}_{{\dot p}{ q}} & 0\cr}
\right)
= - \G^{i~\underline{\a}\underline{\b}}
\end{eqnarray}
($\gamma ^{i}_{q\dot p}$ are $\sigma$--matrices  for $SO(8)$ group,
$\tilde{\gamma }^{i}_{\dot{q} p} \equiv \gamma^{i}_{p\dot q}$),
and the inverse spinor moving frame matrix
\begin{equation}\label{harm1}
v^{~\underline{\mu}}_{\underline{\a}} =
( v^{-\underline{\mu}}_{q}~,~v^{+\underline{\mu}}_{\dot q})
{}~~~~\in ~~~Spin(1,D-1) \qquad
\end{equation}
can not be expressed in terms of the variables \p{harm} in a simple
manner.

For example, for the simplest $D=3$ case the expressions \p{10D},
relating vector
and spinor harmonics, have the form
\begin{eqnarray}\label{3D}
u^{++}_{\underline{ m}}
\Gamma^{\underline{ m}}_{\underline{ \mu} \underline{ \nu}}
= 2 v^{~+}_{\underline{\mu} } v^{~+}_{\underline{\nu} } , \qquad
u^{--}_{\underline{ m}}
\Gamma^{\underline{ m}}_{\underline{ \mu} \underline{ \nu}}
= 2 v^{~-}_{\underline{\mu} }
v^{~-}_{\underline{\nu} } , \qquad \nn
u^{~\perp}_{\underline{ m}}
\Gamma^{\underline{ m}}_{\underline{\mu} \underline{\nu}}
= v^{~+}_{\underline{\mu} } v^{~-}_{\underline{\nu} }
+ v^{~-}_{\underline{\mu} } v^{~+}_{\underline{\nu} }
, \qquad
\end{eqnarray}
with spinor harmonics $v^{\pm}_{\underline{\mu}}$
being bosonic spinors restricted by the normalization conditions
\begin{equation}\label{3Dh}
v^{-\underline{\mu}} v^{+}_{\underline{\mu}} \equiv
\e^{\underline{ \mu} \underline{ \nu}}
v^{-}_{\underline{\nu}}
v^{+}_{\underline{\mu}}
\equiv
v^{-\underline{\mu}}
\e_{\underline{ \mu} \underline{ \nu}}
v^{+}_{\underline{\nu}} = 1 , \qquad
\end{equation}
(named the harmonicity conditions \cite{gikos,sok,bhbz0}) only.

Note that eqs. \p{10D} or \p{3D} result in the orthonormality
relations for the composed moving frame vectors
\begin{equation}\label{o}
u^{~\underline{a}}_{\underline{m}}\eta^{\underline{m}\underline{n}}
u_{\underline{n}\underline{b}} = \eta^{\underline{a}\underline{b}}=
{\it diag}(1,-1,\ldots ,-1)
\end{equation}
which, in particular, include the light--likeness conditions for
$u^{\pm\pm}$:
\begin{eqnarray}\label{o1}
u^{++}_{\underline{m}} u^{++\underline{n}} = ~0~
= u^{--}_{\underline{m}} u^{--\underline{n}}
, \qquad \nn
u^{\pm\pm}_{\underline{m}} u^{\underline{m}i} = 0 , \qquad \nn
u^{++}_{\underline{m}} u^{--\underline{n}} = 2 , \qquad
u^{i}_{\underline{m}} u^{j \underline{m}} = - \d^{ij} , \qquad
\end{eqnarray}
More details about harmonics can be
founded in Refs.  \cite{gikos,sok,bhbz0} - \cite{bzp,bpstv}.

To get the superfield equations of motion
from the generalized action \p{1}, \p{1a}, both the coefficients of
the forms and the bosonic submanifold are varied.

The variation of the
action over ${\cal M}_{2}$ is amount to superdiffeomorphism
transformations on the world supersurface. This allows one to extend the
superfield equations from ${\cal M}_{2}$ to the whole supersurface.

Then we will stress, that intrinsic geometry of the world supersurface
is not {\sl a priori}
restricted by any superfield constraints, and the embedding of the world
supersurface into the target superspace is not {\sl a priori} specified by
any condition such as a geometrodynamical condition
\cite{stv}--\cite{bers94} (see eq. \p{geomd}, the latter playing the
crucial role in the twistor--like superfield approach). All the
constraints  and the geometrodynamical condition are obtained as equations
from the generalized action.

This garantees that the equations of motion, which are the differential
form equations, can be extended to the whole world superspace and that
variations of the integration surface do not give new independent
equations to those which are get by variations of fields.

As we will demonstrate below, the field variations of the action give two
kinds of relations:
\\ $1)$ relations between target superspace and world
supersurface vielbeins which originate them along one another and are the
standard relations of surface embedding theory; we have called
them
''rheotropic'' conditions \cite{bsv}
\footnote{'rheo' is 'current' and 'tropic' is
'direction, rotation' in Greek}; \\ $2)$ dynamical equations causing the
embedding to be minimal.
\\ Only the latter equations put the theory on the
mass shell.

\bigskip

The last term in \p{1a} is the Wess--Zumino 2--form \cite{gsw}.
Its coefficient being fixed by
the requirement that the action \p{1}, \p{1a} has
$(D-2)$-- parametric fermionic gauge symmetry, which is the projection of
the world sheet superspace supertranslations onto an integration surface
${\cal M}^2$.

\bigskip

{}From the rheonomy point of view \cite{rheo}, the theory is off the mass
shell superdiffeomorphism invariant if for the action \p{1} to be
independent of the surface ${\cal M}_{2}$ (i.e. $d{\cal L}_{2} = 0$) only
the rheotropic relations are required, and the latter do not lead to the
equations of motion.

\bigskip

In the next section, after a complete analysis of consequences of
 rheotropic conditions we will prove the off-- shell superdiffeomorphysm
 invariance of the generalized action \p{1}, \p{1a} for heterotic string.

\bigskip

The other evident gauge symmetries of the generalized action \p{1}, \p{1a}
are $SO(1,1)$ (identified with world sheet Lorentz group) and $SO(D-2)$.
This results in the possibility to consider the components of the spinor
moving frame matrix
$v^{~\underline{\mu}}_{\underline{\a}}$
(taking it values in $Spin(1,D-1)$ \p{harm}, \p{harm1}), as well as
the components of orthogonal 'vector' matrix
$u_{\underline{m}}^{\underline{a}}=(u_{\underline{m}}^{{a}},
u_{\underline{m}}^{i}$) (see \p{o}, \p{o1})
as coordinates parametrizing (noncompact)
coset space
${SO(1,D-1) \over {SO(1,1)\times SO(D-2)}}$ (see
\cite{sok,bzst,bzm,bzp,bpstv}).

\subsection{Equations of motion}

Varying the action \p{1}, \p{1a} over $1$--forms $e^{++},~ e^{--}$
and the fields $X^{\underline{m}}$ and $\Th^{\underline{\mu}}$ we
get the following differential form equations
\begin{equation}\label{e}
{\d S\over{\d e^{\pm\pm}}} = 0 ~\Rightarrow~
E^{\mp\mp} \equiv \Pi^{\underline{m}}
u_{\underline{m}}^{\mp\mp} = e^{\mp\mp }
\end{equation}
\begin{eqnarray}\label{x}
{\d S\over{\d X^{\underline m}}} \equiv
{\d S\over{\om^{\underline{m}}(\d)}} = 0
{}~\Rightarrow~  \nn
d( u^{++}_{\underline{m}} e^{--} - u^{--}_{\underline{m}} e^{++})
- 2i d\Th \G_{\underline {m}} d\Th = 0,
\end{eqnarray}
\begin{eqnarray}\label{th}
{\d S\over{\d\Th^{\underline{\mu}}}} \vert_{\om^{\underline{m}}(\d)=0}
{}~~~~ = 0
{}~\Rightarrow~ \nn
d\Th^{\underline{\mu}}\G^{\underline{m}}_{\underline{\mu\nu}}
(u^{--}_{\underline{m}} e^{++} - u^{++}_{\underline{m}} e^{--}
+ 2 \Pi_{\underline{m}}) = 0
\end{eqnarray}

To get the rest of the equations we will perform the varying of the action
with respect to the harmonic variables. The composed nature of
the light--like vectors $u^{\pm\pm }_{\underline{m}}$
(see \p{10D} or \p{3D}
and similar relations for $D=4~and~6$) or, equivalently, the
orthonormality conditions \p{o}, \p{o1} should be taken into account in
such varying.

Indeed, due to the orthonormality conditions \p{o}, the matrix
\begin{equation}\label{harmvec}
||u^{~\underline{a}}_{\underline{m}} ||
\equiv || u^{~a}_{\underline{m}} ~~|~~
u^{i}_{\underline{m}} ||
\equiv || u^{\pm\pm}_{\underline{m}} ~~|~~
u^{i}_{\underline{m}} ||
{}~~~~~~~~\in ~~~~~~~~SO(1,D-1)
\end{equation}
takes its values in the vector representation of the Lorentz group, as well
the matrix $v^{\underline{\a}}_{\underline{\mu}}$ \p{harm} takes its values
in its spinor representation. So, they both have $D(D-1)/2$ degrees of freedom.
Henceforth, the dimension of the space tangent to the harmonic (or moving
frame)
sector should be $D(D-1)/2$ too.

The natural basis for this tangent space is given by
$D(D-1)/2$ Cartan forms
\begin{equation}\label{C}
\Om^{\underline{a}\underline{b}} \equiv
u^{~\underline{a}}_{\underline{m}}
d u^{\underline{b} \underline{m}}
\end{equation}
which split naturally into the set of \\
-- $2(D-2)$ covariant forms
being the basis of the ({\sl noncompact}) coset space
${SO(1,9) \over {SO(1,1) \otimes SO(8)}}$
\begin{equation}\label{10f++}
\Om^{++~i} \equiv u^{++}_{\underline m} d u^{\underline m~i}
= {1 \over 4} v^{+\underline \mu}_{\dot q} \tilde{\g}^{i}_{\dot q q}
dv^{~+}_{\underline \mu q} ,
\end{equation}
\begin{equation}\label{10f--}
\Om^{--~i} \equiv u^{--}_{\underline m} d u^{\underline m~i}
= {1 \over 4} v^{-\underline \mu}_{q} \g^{i}_{q \dot q}
dv^{~-}_{\underline \mu \dot q} ,
\end{equation}
--- $1$ form having the transformation properties of the
$SO(1,1)$ connection
\begin{equation}\label{10f1c}
\Om^{(0)} \equiv {1 \over 2} u^{--}_{\underline m} d u^{\underline m~++}
= {1 \over 4}
v^{-\underline \mu}_{q}
dv^{~+}_{\underline \mu q}
= {1 \over 4}
v^{+\underline \mu}_{\dot q}
dv^{~-}_{\underline \mu \dot q} ,
\end{equation}
and, at least, \\
--  $(D-2)(D-3)/2$ forms being the
$SO(D-2)$   connections
\begin{equation}\label{10f8c}
\Om^{ij} \equiv u^{i}_{\underline m} d u^{\underline m~j}
= - {1 \over 4} v^{-\underline \mu}_{q} {\g}^{ij}_{q p}
dv^{~+}_{\underline \mu p}
= {1\over 4} v^{+\underline \mu}_{\dot q} \tilde{\g}^{ij}_{\dot q \dot p}
dv^{~-}_{\underline \mu \dot p} ,
\end{equation}
(For definiteness, all the expressions in terms of spinor harmonics are
presented for
$D=10$ case).

\bigskip

For the further analyzes it is essential that,
due to \p{C}, $\Om^{\underline{a}\underline{b}}$ satisfy identically
Maurer--Cartan equations
\begin{equation}\label{MC}
d \Om^{\underline{a}\underline{b}} +
\Om^{\underline{a}}_{~\underline{c}}
\Om^{\underline{c}\underline{b}}
= 0
\end{equation}
which split naturally into the set of equations for the forms
\p{10f++}--\p{10f8c}, which are
\begin{equation}\label{PC++}
{\cal D} \Om^{++i} \equiv d  \Om^{++i} -  \Om^{++i} \Om^{(0)}
+  \Om^{++j} \Om^{ji} = 0
\end{equation}
\begin{equation}\label{PC--}
{\cal D} \Om^{--i} \equiv d  \Om^{--i} +  \Om^{--i} \Om^{(0)}
+  \Om^{--j} \Om^{ji} = 0
\end{equation}
\begin{equation}\label{G}
{\cal F} \equiv d \Om^{(0)} = {1\over 2} \Om^{--i} \Om^{++i}
\end{equation}
\begin{equation}\label{R}
R^{ij}  \equiv d \Om^{ij} + \Om^{ik} \Om^{kj} =
- \Om^{--[i} \Om^{++j]}
\end{equation}

The admissible variation of the composed vectors
$u^{~\underline{a}}_{\underline{m}}$ (as well as the variation of the
spinor harmonics $v$)
can be considered as an element of the cotangent space and,
hence, can be decomposed onto the same basis of the forms
(taken to be dependent
on the variation symbol $\d$ instead of the external differential symbol $d$).
So,
\begin{equation}\label{var}
\d u^{\pm\pm}_{\underline{m}} =
\pm 1/2 u^{\pm\pm}_{\underline{m}} \Om^{(0)}(\d) +
u^{i}_{\underline{m}} \Om^{\pm\pm ~i} (\d) ,
\end{equation}
in spite of is the moving frame vectors considered as composed from
the spinor harmonics, or supposed to be fundamental.

Hence, the varying with respect to harmonic variables leads to
the equations
\begin{equation}\label{varom0}
\Sigma_{\pm}
(\pm u^{\mp\mp}_{\underline{m}}
{\d S\over{\d u^{\pm\pm}_{\underline{m}} }})
\equiv {\d S\over{\Om^{(0)}(\d)}} = 0
{}~~~\Rightarrow ~~~ E^{++} e^{--} +  E^{--} e^{++} = 0
\end{equation}
and
\begin{equation}\label{u0}
u^{i}_{\underline{m}} {\d S\over{\d u^{\pm\pm}_{\underline{m}} }}
\equiv
{\d S\over{\Om^{\pm\pm i}(\d)}} = 0
{}~~~\Rightarrow ~~~ E^{i} e^{\mp\mp} = 0
\end{equation}

It is easy to see, that Eq. \p{varom0} is satisfied identically due to \p{e}.
This reflect the local $SO(1,1)$ (world--sheet Lorentz) symmetry of the
considered action.

So, the only independent equations of motion appearing as a result of
the varying with respect to harmonic
(or moving frame) variables is \p{u0}, which
means
\begin{equation}\label{u}
E^{i} \equiv
\Pi^{\underline{m}}
u_{\underline{m}}^{i} = 0
\end{equation}

Eqs. \p{u} and \p{e} are just the bosonic subset of the set of the
rheotropic relations \cite{bsv} for the case of superstring.
They cause the light-- like bosonic components $E^{\pm\pm}$ \p{rep_indb}
of superspace vielbein \p{rep_ind} to become tangent to the world sheet
superspace and the rest of them to become orthogonal one.

So, the bosonic part of rheotropic conditions can be rewritten
as unique vector $1$--form equation
\begin{equation}\label{pi}
\Pi^{\underline{m}} = {1 \over 2} (e^{++} u^{--}_{\underline{m}} + e^{--}
u^{++}_{\underline{m}} )
\end{equation}
which is the supersymmetric
counterpart of the basic relations of the geometric approach to bosonic
string theory \cite{geo}.  But here, following \cite{bsv}, we have
derived it from the action principle for superfield case \footnote{See the
first section of \cite{bpstv} for the similar result for the bosonic
string.  Such component (not superfield) supersymmetric equations had been
derived from the action principle in refs.  \cite{bzst,bzm,bzp}, however
without discussion of the relation with the geometric approach}).

Moreover, now Eq.\p{pi} has the projection onto the Grassmann directions
of the world sheet tangent superspace
\begin{equation}\label{geomd}
\Pi^{\underline{m}}_{+q} = 0
\end{equation}
which is just the Geometrodynamic equation \cite{stv}--\cite{bers94},
which was the
basis of the previous doubly supersymmetric formulations.

Substituting \p{pi} into the equations \p{th} we derive the simple two
form equation
\begin{equation}\label{th1}
e^{++} (d\Th \Gamma)_{\underline{\mu}} u^{--}_{\underline{m}}
=0,
\end{equation}
which, using for the $u^{--}_{\underline{m}}$ the expression from the
first line of Eq. \p{10D} or \p{3D}, can be further simplified
\begin{equation}\label{th2}
e^{++} d\Th^{\underline{\mu}} v_{\underline{\mu} \dot q}^{~-} = 0,
\end{equation}
(for $D=3$ the symbol $\dot q$ should be omitted). The latter is most
suitable for the further analyzes.

Below, after the consideration of the relation with the component
 twistor--like superstring formulation of Refs. \cite{bzst,bzm,bzp} we will
 study the set of equations \p{u}, \p{e} (or \p{pi}), \p{th2}, \p{x} for
 the case of $D=10$ heterotic superstrings
 (the results for simpler cases
 $D=3, 4 ~and~ 6$ can be derived by reduction).

 It will be proved that Eqs. \p{x} are always dependent on \p{th2}, \p{u}
 and \p{e}. Moreover, we will prove that the only dynamical equation in
 the set of the rest relations are the component of \p{th2} appeared as
 the coefficient for the basic two--form $e^{++}~e^{--}$ and having the
 form
\begin{equation}\label{th1c}
\nabla_{--} \Th^{\underline{\mu}}
v^{~-}_{\underline{\mu} \dot q} = 0 ,
\end{equation}

Eq. \p{th1c} coincides formally with the equation of motion for the
field $\th$ appearing in the component twistor--like formulation
of Refs. \cite{bzst,bzm,bzp}. The another equation contained in \p{th2}
\begin{equation}\label{th1s}
\nabla_{+ q} \Th^{\underline{\mu}} v^{~-}_{\underline{\mu} \dot q} = 0,
\end{equation}
can be considered as the fermionic part of the set of the rheotropic
conditions and, as it will be proved below, do not lead to any dynamical
equations.

This will be done by the investigation of the selfconsistency conditions
for these equations.

The presence of the spinor moving frame variables
(Lorentz harmonics) gives, from one hand, the possibility to formulate the
doubly supersymmetric geometrical approach \cite{bpstv}
to heterotic superstring as the
result of such investigation, and require, from the other hand, to
investigate the Maurer--Cartan equations \p{MC} (or Eqs.
\p{PC++} -- \p{R} being the counterpart of
the Peterson--Codazzi, Gauss and Rici equations) as the part of the
selfconsistency conditions.

Moreover, we will prove that
the closure of the Lagrangian $2$--form
holds when only the equations \p{u}, \p{e} (or \p{pi}) and the
equation \p{th1s} are taken into account.
This reflect the off --shell diffeomorphism invariance of the discussed
action for $D=3,~4, ~6~and ~10$ heterotic string in the rheonomic sense.

\subsection{Component formulation and local fermionic symmetry}

The component formulation \cite{bzst,bzm,bzp} of the heterotic
superstring is obtained
by choosing the surface ${\cal M}_{2}$ to be defined by the condition
$\eta^{+q}=0$ and taking  into account only the vector components of
\p{d}. In this case the action \p{1} is just the action of Refs.
\cite{bzst,bzm,bzp} for the fields
$$X^{\underline m}|_{\eta=0}=x^{\underline m}(\xi), \qquad
\Th^{\underline \mu}|_{\eta=0}=\th^{\underline \mu}(\xi), \qquad $$
$$u^{\underline{m}}_a|_{\eta=0}=u^{\underline{m}}_a(\xi) \qquad  and
\qquad  e^a|_{\eta=0}=e^a(\xi) .$$
but rewritten in terms of the differential forms.

For these fields one can get from \p{u}--\p{th1} the
following equations:
\begin{equation}\label{pic}
\Pi_{\pm\pm}^{\underline{m}}=e_{\pm\pm}^m(\partial_m x^{\underline{m}}
-i\partial_m\th\G^{\underline{m}}\th)=
u^{\underline{m}}_{\mp\mp}(\xi),
\end{equation}
\begin{equation}\label{thc}
e^{m}_{--} \partial_m \th^{\underline{\mu}}
v^{~-}_{\underline{\mu} \dot q} = 0 ,
\end{equation}
\begin{equation}\label{xc}
\partial_m (e e^{m}_{\mp\mp} u^{\pm\pm\underline m}) -
4i \varepsilon^{m n}
\partial_{m}\th \G^{\underline{m}} \partial_{n}\th = 0,
\end{equation}
Which are just the string equations of the component twistor--like
formulation  \cite{bzst,bzm,bzp}.

Using the component rheotropic equation \p{pic} (which can be transformed
into the form representing any of the sets of the vector variables
$\Pi_{m}^{\underline{m}}, ~u^{\pm\pm}_{\underline{m}}, ~or~ e^{\pm\pm}_{m}$
through two others) we can transform equations \p{thc} and \p{xc} into the
form of the standard Green--Schwarz formulation \cite{gs,gsw}
\begin{equation}\label{thc0}
\Pi^{\underline{m}}_m g^{mn}
\partial_n \th^{\underline{\mu}}
\G^{~\underline{\mu} \underline{\nu}}_{\underline{m}}
= 0 ,
\end{equation}
and
\begin{equation}\label{xc0}
\partial_m (\sqrt{-g}g^{mn}\Pi^{\underline m}_n)
- 2i \varepsilon^{m n}
\partial_{m}\th \G^{\underline{m}} \partial_{n}\th = 0,
\end{equation}
where $g_{mn}=e_m^ae_{an}= \Pi^{\underline m}_m \Pi_{{\underline m}n}$
is the induced metric on the world sheet.

The component action obtained from \p{1}, \p{1a} by choosing
${\cal M}^2 =  {\cal M}^2_0$ \p{m20}
possess the $\k$-symmetry
in the following irreducible form
\cite{bzst,bzm,bzp}
\begin{eqnarray}\label{k}
\d \th^{\underline{\mu}} = \k^{+q} v^{-\underline \mu}_{q} \qquad \nn
\omega^{\underline m} (\d) = 0 , \qquad \Rightarrow \qquad
\d x^{\underline{m}} = i \k^{+q} v^{-\underline \mu}_{q}
   \G^{\underline m}_{\underline{\mu}\underline{\nu}}
\th^{\underline{\mu}} \qquad \nn
\d e^{++} = - 4i (d\Th v^{+}_{q}) \k^{+q} , \qquad \d e^{--} = 0 ,
\qquad
\end{eqnarray}

\begin{eqnarray}\label{dv}
\d v^{~+}_{\underline{\mu} q} = 2 i \kappa^{+p} \gamma^{i}_{p \dot p}
e^{m}_{--} \partial_m \th^{\underline{\nu}}
v^{~-}_{\underline{\nu} \dot p}
\gamma^{i}_{q \dot q} v^{~-}_{\underline{\mu} \dot q} , \qquad \nn
\d v^{~-}_{\underline{\mu} \dot q} = - 2 i \kappa^{+p}
\gamma^{i}_{p \dot p} e^{m}_{++} \partial_m \th^{\underline{\nu}}
v^{~-}_{\underline{\nu} \dot p}
\gamma^{i}_{q \dot q} v^{~+}_{\underline{\mu} q} ,
\end{eqnarray}

The basic feature of the twistor--like superfield approach is that this
transformations \p{k}, \p{dv} are the relic of the world surface
superdiffeomorphisms \cite{stv}, for instance, $\th^{\underline\mu}$ and
$v^{\underline\mu}_{\a p}$ are transformed as superpartners.

\bigskip

However, if we consider  world surface superspace diffeomorphysms as the
symmetry of the generalized action \p{1}, \p{1a}, they will be projected
onto an integration surface ${\cal M}^2$ and, so, are realized
nonlinearly.

 Namely, the fermionic symmetry of the generalized action \p{1}, \p{1a} is
 defined by relations \p{k} with all the fields
replaced by superfields taken at a surface ${\cal M}^2$ \p{m2} and
variations of harmonic superfields are defined by relations
 \begin{equation}\label{dvs}
 \d v^{~+}_{\underline{\mu} q} = 1/2 \Om^{++i}(\d ) \g^i_{q\dot q}
 v^{~-}_{\underline{\mu} \dot q}  \qquad
 \d v^{~-}_{\underline{\mu} \dot q} = 1/2 \Om^{--i}(\d )
 v^{~+}_{\underline{\mu}  q}
 \g^i_{q\dot q}
 \end{equation}
 with $\Om^{\pm\pm}(\d )$ being determined by the solution of 1--form
 equation
 \begin{equation}\label{ksf}
 e^{--} \Om^{++i}(\d )  -  e^{++}
 \Om^{--i}(\d ) + 4i d\Th v^{-}_{\dot q} \g^i_{q\dot q} \kappa^{+}_{q} = 0
 \end{equation}
 In \p{ksf} all the forms are pulled back on the surface
 ${\cal M}^2$ \p{m2}, i.e.
 \begin{equation}\label{pullback}
 e^{--} = d\xi^m e_{m}^{--} (\xi, \eta (\xi)) + d\eta^q (\xi)
 e_q^{--}(\xi, \eta (\xi)) = d\xi^m (e_{m}^{--} (\xi, \eta (\xi)) + \partial_m
 \eta^q (\xi) e_q^{--}(\xi, \eta (\xi))
 \end{equation}
 ets.

 For the choice of surface ${\cal M}^2 = {\cal M}^2$, which corresponds
to the component formulation, the solutions of Eq. \p{ksf} define just the
transformation rules \p{dv} for harmonic variables.

\section{$D = 10$ heterotic superstring:
Doubly supersymmetric geometric approach
 and the 'off--shell' superdiffeomorphism invariance of
 the generalized action.}

In this section we will investigate completely the set of the superfield
equations following from the generalized action principle \p{1} for
$D=10$ heterotic string.
Such investigation naturally results in the construction of the
doubly supersymmetric geometric approach \cite{bpstv}
for the heterotic string in $D=10$
\footnote{but based on the generalized action principle instead of the
geometrodynamic equation, as it was in \cite{bpstv}; it should be stressed,
that just for the case of heterotic string, where the
Geometrodynamic condition
\p{geomd} do not lead to the equations of motion and, moreover, the
complete superfield form of the equations of motion had not be known,
it was not completely understood
previously how to formulate the minimal embedding of
the heterotic string}.

We will prove that the set of the rheotropic equations \p{u}, \p{e} and
\p{th1s} do not result in any dynamical equation. This will be done by
studying of all the consequences of the rheotropic equations; as a result
we will construct the geometric approach based on the rheotropic equations
only and prove that they define the nonminimal embedding of the world
sheet into the target superspace. We will prove also that the equations
\p{x} are always dependent on the other ones and that the only
independent dynamical equation for the case of the
heterotic string is \p{th1c}.

At the end of the section we present also the
complete set of the equations of the geometrical approach, which
describes the minimal embedding of the heterotic string
into the $D=10$ target superspace.
The results for $D=3, 4 ~and~ 6$ can be
derived by reduction.

Such set of the  equations for $D=3$ heterotic string will be used in the
next section for the investigation of the relation between the heterotic
string and the  supersymmetric extension of the nonlinear Liouville
equation.

\subsection{The conventional rheotropic conditions and the choice of the
world--sheet connections}

\subsection{The geometry of the world sheet superspace}

The world--sheet geometry
(or the world--sheet supergravity)
can be described by the
\begin{eqnarray}\label{wsforms}
supervielbein~~forms: \qquad
e^{A}(d) \equiv (e^{\pm \pm}, e^{+q}) ,
\qquad \nn
Lorentz~~(or~~SO(1,1))~~
connection~~form: \qquad w(d) = e^{A}(d) w_{A} ,
\qquad  \nn
and      \qquad \nn
SO(D-2)~~
connection~~form: \qquad B^{ij}(d) = e^{A}(d) B^{ij}_{A} ,
\qquad  \nn
\end{eqnarray}

However, only the bosonic vielbein forms $e^{\pm \pm}$ are involved
explicitly into the action \p{1},\p{1a}.

Hence, we can choose arbitrary the
 Grassmann vielbein form $e^{+q}$ and
the connections of the both types  $w(d)$ and $B^{ij}(d)$.

The most natural way is to choose
them being induced by the embedding.

For the connections this means, that they are chosen
to be equal to the pull-- backs of the corresponding
Cartan forms \p{10f1c} and \p{10f8c}.
Such coincidence can be formulated as
\begin{equation}\label{10wsc}
\Om^{(0)}({\cal D}) \equiv
 \Om^{(0)}(d) - 2w = 0  , \qquad
\Om^{ij}({\cal D})  \equiv
\Om^{ij}(d) - B^{ij}  = 0  . \qquad
\end{equation}
where ${\cal D}$ is differential covariant with respect  to both world
sheet Lorentz ($SO(1,1)$) and natural 'internal' $SO(D-2)$ symmetries.

As the result of \p{10wsc}, the covariant world surface derivatives
of the spinor and vector moving frame variables acquire the forms
\begin{equation}\label{10dv}
{\cal D} v^{~+}_{\underline \mu q} = {1 \over 2}
\Om^{++~i}(d) \g^{i}_{q \dot q} v^{~-}_{\underline \mu \dot q}  , \qquad
{\cal D} v^{~-}_{\underline \mu \dot q} = {1 \over 2}
v^{~+}_{\underline \mu q} \g^{i}_{q \dot q}
\Om^{--~i}(d) , \qquad
\end{equation}
\begin{equation}\label{10dvi}
{\cal D} v^{~-\underline \mu}_{ q} = - {1 \over 2}
\Om^{--~i}(d) \g^{i}_{q \dot q} v^{~+\underline \mu}_{\dot q}  , \qquad
{\cal D} v^{~+\underline \mu}_{\dot q} = - {1 \over 2}
v^{~-\underline \mu}_{q}  \g^{i}_{q \dot q}
\Om^{++~i}(d) , \qquad
\end{equation}
and
\begin{equation}\label{10du}
{\cal D} u^{++}_{\underline m} =
u^{i}_{\underline m} \Om^{++~i}(d) , \qquad
{\cal D} u^{--}_{\underline m} = u^{i}_{\underline m}
\Om^{--~i}(d) , \qquad
{\cal D} u^{i}_{\underline m} =
u^{--}_{\underline m} \Om^{++~i}(d) +
u^{++}_{\underline m} \Om^{--~i}(d) , \qquad
\end{equation}
respectively.

The fermionic vielbein induced by the embedding has the form
\begin{equation}\label{rh4}
e^{+q} = E^{+q} \equiv
d\Theta^{\underline{\mu}} v^{~+}_{\underline{\mu} q} ,
\end{equation}
 Eq. \p{rh4} is the evident Grassmann counterpart of the rheotropic
 conditions \p{e}. However it is not derived as an equation of motion,
 but chosen using the arbitrariness or symmetry of the action
 \footnote{Let us stress that they can be discussed as the result of the
gauge fixing for the redefinition ''symmetry'' $$ e^{+q} \longmapsto
\tilde{e}^{+q} = (e^{+p}  + e^{\pm\pm} \chi^{~~+p}_{\pm\pm}) W^{~q}_{p}
\qquad  det W \not= 0 $$ which holds due to the mentioned absence of the
Grassmann vielbein in the action \p{1} in the proper form (i.e. it is
present in the external differential decomposition \p{d} only).

The related redefinition of the derivatives (which follows from
$d \longmapsto d$ and $e^{\pm\pm} \longmapsto e^{\pm\pm}$) is
$$
\nabla_{+q} \longmapsto  \tilde{\nabla}_{+q} = (W^{-1})^{~p}_{q}
\nabla_{+p} \qquad \nabla_{\pm\pm} \longmapsto \tilde{\nabla}_{\pm\pm} =
\nabla_{\pm\pm} - \chi^{~~+p}_{\pm\pm} \nabla_{+p} , \qquad $$ }.

So, it is
naturally to refer on it  as on the {\sl \large conventional rheotropic
condition}.

Other rheotropic conditions are \p{e}, \p{u}
\begin{equation}\label{rh1}
E^{++} \equiv \Pi^{\underline m} u^{++}_{\underline m} =
e^{++} ,
\end{equation}
\begin{equation}\label{rh2}
E^{--} \equiv \Pi^{\underline m} u^{--}_{\underline m} =
e^{--} ,
\end{equation}
\begin{equation}\label{rh3}
E^{i} \equiv \Pi^{\underline m} u^{i}_{\underline m} =
0 ,
\end{equation}
and \p{th1s}. The
latter can be presented in terms of 1--forms as follows
\begin{equation}\label{rh5}
E^{-\dot q} = d\Theta^{\underline{\mu}} v^{~-}_{\underline{\mu} \dot q} =
e^{\pm\pm} \psi^{~~~-}_{\pm\pm \dot q} ,
\end{equation}

\subsection{The inducing of the torsion constraints
and the doubly SUSY geometric approach generation}

Let us investigate the selfconsistency conditions for the rheotropic relations
\p{rh4} -- \p{rh5}.

Selfconsistency conditions for eq. \p{rh1}, after taking into
account \p{rh3}, acquire the form
$$
T^{++} \equiv {\cal D} e^{++} = -2i d\Theta v^{+q} d\Theta v^{+q}
$$
which, after taking into account the conventional rheotropic condition
\p{rh4}, coincides with the component of the
flat torsion of the world sheet superspace
\begin{equation}\label{T++}
T^{++} \equiv {\cal D} e^{++} = -2i e^{+q} e^{+q}
\end{equation}

The selfconsistency conditions for \p{rh2}
after taking into account \p{rh3} and \p{rh5}
acquire, respectively, the forms
$$
T^{--} \equiv {\cal D} e^{--} =
-2i d\Theta v^{-}_{\dot q} d\Theta v^{-}_{\dot q}
$$
and
\begin{equation}\label{T--}
T^{--} \equiv {\cal D} e^{--} =
- 4i e^{++} e^{--} \psi^{~~~-}_{++\dot q}
\psi^{~~~-}_{--\dot q}  .
\end{equation}

So the torsion constraints of the 'heterotic' supergravity
\cite{to} are reproduced now as the selfconsistency conditions of the
''tangent vector'' rheotropic conditions \p{rh1}, \p{rh2}.

The selfconsistency conditions for ''orthogonal vector'' rheotropic relation
\p{rh3} produce the following conditions for the forms
$\Om^{++i}$ and $\Om^{--i}$
\begin{equation}\label{sc3}
(\Omega^{++i} + 4i e^{+q} \psi^{~~~-}_{--\dot q}) e^{--} +
(\Omega^{--i} + 4i e^{+q} \psi^{~~~-}_{++\dot q}) e^{++}
= 0 ,
\end{equation}
This means, in particular, that their bosonic components
with zero $SO(1,1)$ weight
coincides and are equal to the main curvatures $h^i$ of the embedded
(super)surface
\begin{equation}\label{hi}
\Omega^{~++i}_{++} = \Omega^{~--i}_{--} = h^i ,
\end{equation}
(see \cite{bpstv} and refs. therein).
Eq. \p{sc3} contain also the expressions for the spinor components
of the Cartan forms through the superfields $\psi$
\begin{equation}\label{omf}
\Omega^{++i}_{+q} = - 4i \g^i_{q\dot q} \psi^{~~~-}_{--\dot q} ,
 \qquad \Omega^{--i}_{+q} = - 4i  \g^i_{q\dot q}
 \psi^{~~~-}_{++\dot q} , \qquad
 \end{equation}
So the only components of the forms $\Om^{\pm\pm i}$
undetermined by \p{sc3} are bosonic ones with the Weyl weights $\pm 4$, namely
$\Om^{~\pm\pm i}_{\mp\mp}$. Hence
\begin{equation}\label{om--}
\Omega^{--i} = - 4i e^{+q} \g^i_{q\dot q} \psi^{~~~-}_{++\dot q} +
e^{++} \Omega^{~--i}_{++} + e^{--} h^i ,
\end{equation}
\begin{equation}\label{om++}
\Omega^{++i} = - 4i e^{+q} \g^i_{q\dot q} \psi^{~~~-}_{--\dot q}
+ e^{--} h^{i} + e^{--} \Omega^{~++i}_{--}
\end{equation}

Then, the selfconsistency conditions for the conventional rheotropic
equation \p{rh4} result in the expression for the spinor torsion
$T^{+q} \equiv {\cal D} e^{+q}$ of the world--sheet superspace
\begin{equation}\label{Tf}
T^{+q} \equiv {\cal D} e^{+q} =
- 2i e^{\pm\pm} e^{+p} \g^i_{p \dot p}
\g^i_{q \dot q} \psi^{~~~-}_{--\dot q}
\psi^{~~~-}_{\pm\pm\dot q} +
 e^{++} e^{--} {1 \over 2}
 (\Om^{~++i}_{--} \g^i_{q \dot q} \psi^{~~~-}_{++\dot q} -
h^i \g^i_{q \dot q} \psi^{~~~-}_{--\dot q}) ,
\end{equation}
which means, in particular,
that the well known conventional torsion constraint holds
$$
T^{~~~~+q}_{+p~+r} = 0 .
$$

Hence we reproduce (as the result of the selfconsistency  of the
rheotropic conditions) the ordinary form of the
covariant spinor derivative algebra
\begin{equation}\label{der}
\{ {\cal D}_{+q}, {\cal D}_{+p}\} = 4i \d_{qp} {\cal D}_{++} +
{}~~curvature~~
\end{equation}

The selfconsistency conditions for the last rheotropic equation \p{rh5}
produce the restrictions on the $\psi$ superfields
\begin{equation}\label{r++}
{\cal D}_{+q} \psi^{~~~-}_{++\dot q} = - {1 \over 2} \g^i_{q\dot q}
\Omega^{~--i}_{++}
 \end{equation}
 \begin{equation}\label{r--}
{\cal D}_{+q} \psi^{~~~-}_{--\dot q} = - {1 \over 2} \g^i_{q\dot q} h^i
 \end{equation}
\begin{equation}\label{r++--}
{\cal D}_{--} \psi^{~~~-}_{++\dot q} =
{\cal D}_{++} \psi^{~~~-}_{--\dot q} + 4i
\psi^{~~~-}_{++\dot p} \psi^{~~~-}_{--\dot p} \psi^{~~~-}_{--\dot q}
 \end{equation}

The only nontrivial component of the Peterson--Codazzi equations
is one of ${\cal D} \Om^{++i}$, which are proportional to the
basic two form $e^{--} e^{+q}$:
\begin{equation}\label{dom}
{\cal D}_{+q} \Om^{~++i}_{--} = - 4i
(\g^i_{q \dot q}
{\cal D}_{--} \psi^{~~~-}_{--\dot q}
- 2 \g^j_{q \dot q} \g^j_{r \dot p} \g^i_{r \dot r}
\psi^{~~~-}_{--\dot q} \psi^{~~~-}_{--\dot p} \psi^{~~~-}_{--\dot r})
 \end{equation}

The Gauss and Ricci equation define the $SO(1,1)$ and $SO(8)$
curvatures  ${\cal F}$ and $F^{ij}$, which satisfy the Bianchi
identities \begin{equation}\label{BidTf} {\cal D} T^{+q} = - {1 \over 2}
 e^{+q} {\cal F} + {1 \over 4} e^{+p} \g^{ij}_{qp} F^{ij} \end{equation}
\begin{equation}\label{BidT++}
 {\cal D} T^{++} = - e^{++} {\cal F}
  \end{equation}
\begin{equation}\label{BidT--}
 {\cal D} T^{--} = e^{--} {\cal F}
  \end{equation}

\subsection{Minimal embedding of the
 heterotic string world sheet superspace. }

As it can be seen from eq. \p{r--}, the only independent superfield
dynamical equation of motion is just eq. \p{th1c}
\begin{equation}\label{eqm}
 \psi^{~~~-}_{--\dot q} \equiv
\nabla_{--} \Theta^{\underline \mu} v^{~-}_{\underline{\mu}\dot q} = 0
  \end{equation}
which  means in particular the vanishing of the main curvatures
\begin{equation}\label{minbose}
h^i = 0
  \end{equation}
and, hence, the minimality of the embedding \cite{bpstv}.

\bigskip

Projecting equations of motion \p{x} for $X$ superfield onto different
components of vector moving frame \p{harmvec}, it  is easy to see that all
of them are satisfied identically due to rheotropic conditions \p{rh4}
-- \p{rh3} and Eq.\p{eqm}.

So, the equations for  $X$ superfield are dependent ones.

\bigskip

Hence, the minimal embedding of heterotic superstring
world sheet superspace into a flat $D=10~~(and~~3, 4, 6)$ target
superspace is described in terms of
\\ {\bf --i--} the matter superfields
$\psi^{~~~-}_{++\dot q}$ and $\Om^{~++i}_{--}$, which satisfy the
equations
\begin{equation}\label{d--psi}
{\cal D}_{--} \psi^{~~~-}_{++\dot q} = 0 ,
  \end{equation}
\begin{equation}\label{d+psi}
{\cal D}_{+q} \psi^{~~~-}_{++\dot q} = - {1 \over 2} \g^i_{q \dot q}
\Om^{~--i}_{++} ,
  \end{equation}
(where $\Om^{~--i}_{++}$ is difined just by this condition),
\begin{equation}\label{d+om}
{\cal D}_{+q} \Om^{~++i}_{--} = 0 ,
  \end{equation}
and \\
{\bf --ii--} vielbeins $e^A \equiv (e^{\pm\pm}, e^{+q})$
(world--sheet ''supergravity''), restricted by the torsion constraints
\begin{equation}\label{T++0}
T^{++} \equiv {\cal D} e^{++} = -2i e^{+q} e^{+q}
\end{equation}
\begin{equation}\label{T--0}
T^{--} \equiv {\cal D} e^{--} = 0
\end{equation}
\begin{equation}\label{Tf0}
T^{+q} \equiv {\cal D} e^{+q} =
 e^{++} e^{--} {1 \over 2}
\Om^{~++i}_{--} \g^i_{q \dot q} \psi^{~~~-}_{++\dot q}
\end{equation}

In fact, the matter superfields define the torsion components except
for one leaving nonvanishing in the flat limit.

The curvatures are defined by Gauss and Ricci equations and have the forms
$$SO(1,1)$$
\begin{equation}\label{so(1,1)}
{\cal F} \equiv d\Om^{(0)} =
2i e^{--} e^{+q}  \g^i_{q \dot q}  \Om^{~++i}_{--}
\psi^{~~~-}_{++\dot q} + {1 \over 2}
 e^{++} e^{--} \Om^{~--i}_{++} \Om^{~++i}_{--}
\end{equation}
and, for $D=4, 6, 10$,
$$ SO(D-2) $$
\begin{equation}\label{so(8)}
 F^{ij} =
4i e^{--} e^{+q} \Om^{~++[i}_{--} \g^{i]}_{q \dot q}
\psi^{~~~-}_{++\dot q}
- e^{++} e^{--} \Om^{~--[i|}_{++} \Om^{~++|j]}_{--}
\end{equation}

\subsection{The action independence on the surface}

The conditions of the off--shell superdiffeomorphysm invariance of the
action \p{1} for $D=10~~(and~ 3, 4, 6)$ heterotic string has the form
\begin{eqnarray}\label{dL}
d{\cal L}_2 = {1\over 2 } [ &
(E^{++} - e^{++}) (T^{--}
- 2i d\Theta v^{-}_{\dot q}~d\Theta v^{-}_{\dot q}) - \nn &
(E^{--} - e^{--}) (T^{++}
- 2i d\Theta v^{+}_q ~d\Theta v^{+}_q ) - \nn &
E^{i}  (\Omega^{++i} e^{--} - \Omega^{--i} e^{++}  -
4i d\Theta v^{+}_q~ \g^i_{q\dot q}d\Theta v^{-}_{\dot q}) - \nn &
- 4i e^{++} d\Theta v^{-}_{\dot q}~d\Theta v^{-}_{\dot q}) ]
= 0
\end{eqnarray}
The first three terms vanish due to the rheotropic equations \p{u} and
\p{e}.
The last term (after the complete decomposition onto the supervielbein
forms) can be rewritten as follows
$$
- 4i e^{++} e^{+q}
[2 e^{--} \nabla_{--}\Theta v^{-}_{\dot q} + e^{+p} \nabla_{+p} \Theta
v^{-}_{\dot q} ] ~\nabla_{+q}\Theta v^{-}_{\dot q} $$
It is evident, that
the latter expression vanishes due to the ''fermionic'' rheotropic
equation \p{th1s} only.

Hence,
$$ d{\cal L}_2 = 0 $$
holds as the result of the rheotropic conditions only.

This means the off--shell superdiffeomorphism invariance of the heterotic
 string action in the rheonomy sense \cite{bsv}), because it have been
 proved above that the rheotropic conditions do not lead to the equations
 of motion.

 \section{$D=3$ heterotic superstring and $n=(1,0)$ supersymmetric
 generalization of nonlinear Liouville equation}

In conclusion, let us analyze the set of geometric approach equations
\p{d--psi} -- \p{so(1,1)} for the simplest case of $D=3, N=1$ string
 (where Eq. \p{so(8)} is absent).

Eq.\p{d+om} and the consequence
$$
{\cal D}_{+} {\cal D}_{--} \psi^{~~-}_{++} =
{\cal D}_{--} {\cal D}_{+} \psi^{~~-}_{++} = 0
$$
of Eq. \p{d--psi} can be used to determine $SO(1,1)$ connection
\begin{equation}\label{1NOm0}
\Om^{(0)} =
{1\over {2 \Om^{~++}_{--} }}
(e^{+} \nabla_+ + e^{++} \nabla_{++}) \Om^{~++}_{--}
-
{1\over {2 {\cal D}_{+} \psi^{~~-}_{++}  }}
e^{--} \nabla_{--} {\cal D}_{+} \psi^{~~-}_{++}  ,
\end{equation}
Then Eq. \p{so(1,1)} reproduces the equation
\begin{equation}\label{sleqcov}
{\cal D}_{+} \nabla_{--}
lg(\Om^{~++}_{--} {\cal D}_{+} \psi^{~~-}_{++} ) =
- 4i \Om^{~++}_{--} \psi^{~~-}_{++}
\end{equation}
and its consequence
\begin{equation}\label{cleqcov}
{\cal D}_{++} \nabla_{--}
lg|\Om^{~++}_{--} {\cal D}_{+} \psi^{~~-}_{++} | =
- 2 \Om^{~++}_{--} {\cal D}_{+} \psi^{~~-}_{++}
\end{equation}

Taking into account \p{1NOm0}, we can rewrite Eqs. \p{T--0} -- \p{Tf0} as
follows
\begin{equation}\label{clos--}
d((\Om^{~++}_{--})^{1/2} e^{--}) = 0
\end{equation}
\begin{equation}\label{clos++}
d(( {\cal D}_{+} \psi^{~~-}_{++})^{1/2} e^{++}) = - 2i
\tilde{e}^+ \tilde{e}^+
\end{equation}
\begin{equation}\label{clos+}
d\tilde{e}^+ = 0
\end{equation}
where
\begin{equation}\label{tae+}
\tilde{e}^+ \equiv
({\cal D}_{+} \psi^{~~-}_{++})^{1/4}
(e^+ - i/8 e^{++} \nabla_{+}
lg|\Om^{~++}_{--} {\cal D}_{+} \psi^{~~-}_{++}|
\end{equation}
In the derivation of Eq.\p{clos+} ~~~~ Eq. \p{sleqcov} should be taken into
account.

Relations \p{clos--} -- \p{clos+} coincides with the expressions for the
torsion of flat superspace. This reflects the known statement: two
dimensional $n=(1,0)$ supergeometry is always conformally flat
\cite{nilmo}.

Hence, neglecting possible inputs from a nontrivial world sheet topology,
we can represent supervielbein as follows
\begin{equation}\label{ex--}
e^{--} = (\Om^{~++}_{--})^{-1/2} d\xi^{(--)}   ,
\end{equation}
\begin{equation}\label{ex++}
e^{++} =
({\cal D}_{+} \psi^{~~-}_{++})^{-1/2} w^{(++)}
\equiv
({\cal D}_{+} \psi^{~~-}_{++})^{-1/2} (d\xi^{(++)}
- 2i d\eta^+ \eta^+ ) ,
\end{equation}
\begin{equation}\label{ex+}
e^+ =
({\cal D}_{+} \psi^{~~-}_{++})^{- 1/4} (d\eta^+
+ i/8 w^{(++)} D_{(+)}
lg|\Om^{~++}_{--} {\cal D}_{+} \psi^{~~-}_{++}| ,
\end{equation}
where
$$
w^{(++)} \equiv  d\xi^{(++)}
- 2i d\eta^+ \eta^+  ,
$$
is the basic supersymmetric  1--form
\cite{volkov} of a flat $d=2, n=(1,0)$ superspace and
\begin{equation}\label{derfl}
D_{(+)} \equiv {\partial \over {\partial \eta^{(+)} } } +
2i \eta^+ \partial_{(++)}
\end{equation}
is flat covariant derivative of the superspace.

It is convenient to identify variables $\xi^{(\pm\pm)},~ \eta^{(+)}$ with
local coordinates of world sheet superspace. Of course, the gauge with
respect to superdiffeomorphisms  is fixed by this step. Only $SO(1,1)$
gauge invariance remains unbroken.

Now we are ready to present geometric approach equations in terms of flat
superspace derivatives $\partial_{(\pm\pm)} = \partial /
\partial_{(\pm\pm)}$ and $D_{(+)}$ \p{derfl}.

First of all, let us note, that Eq.\p{d--psi}
${\cal D}_{--} \psi^{~~-}_{++}= 0$ can be represented as a flat space
chirality condition
\begin{equation}\label{Psich}
\partial_{(--)} \Psi^{(+)}_L = 0
\end{equation}
for fermionic superfield
\begin{equation}\label{Psi}
 \Psi^{(+)}_L =
 ({\cal D}_{+} \psi^{~~-}_{++})^{-3/4}  \psi^{~~-}_{++}
\end{equation}

Now, if we try to write the expression
 ${\cal D}_{+} \psi^{~~-}_{++}$ in terms of superfield
  $\Psi^{(+)}_{L}$ and flat fermionic derivative \p{derfl}, we get the
  following identity
  \begin{equation}\label{ident}
  (\Om^{~++}_{--} {\cal D}_{+} \psi^{~~-}_{++})^{3/4}
  \equiv D_{(+)} ((\Om^{~++}_{--} {\cal D}_{+} \psi^{~~-}_{++})^{3/4}
  \Psi^{(+)}_L )
  \end{equation}

  Let us introduce the gauge invariant superfield $W$ by
  \begin{equation}\label{W}
  exp\{ 4W\} = \Om^{~++}_{--} {\cal D}_{+} \psi^{~~-}_{++}
  \end{equation}
  Then the identity \p{ident} gives us the connection between
  $W$ and $\Psi^{(+)}_L$
  \begin{equation}\label{ident1}
  D_{(+)} (e^{3W} \Psi^{(+)}_L ) = e^{3W} ~~~~~~
  \Leftrightarrow  ~~~~~
  D_{(+)} \Psi^{(+)}_L = 1 - 3 D_{(+)} W \Psi^{(+)}_L
\end{equation}

 \bigskip

 Now all the supervielbeins and Cartan forms can be expressed in terms of
 superfields  $\Psi^{(+)}_L, ~W$ and
 \begin{equation}\label{L}
  L = 1/4 lg |{\cal D}_{+} \psi^{~~-}_{++} / \Om^{~++}_{--}|
 \end{equation}
 (being the compensator for $SO(1,1)$ gauge transformations) as follows
\begin{equation}\label{1NOm--f}
\Om^{--} = e^{W+L} ( - 2 w^{(++)} (1 -  D_{(+)} W \Psi^{(+)}_L) -
4i d\eta^{(+)} \Psi^{(+)}_L ) ,
\end{equation}
\begin{equation}\label{1NOm++f}
\Om^{++} =  d\xi^{(--)} e^{W+L}
\end{equation}
\begin{equation}\label{1NOm0f}
\Om^{(0)} =
(d\eta^{(+)} D_{(+)} + w^{(++)} \partial_{(++)}
-  d\xi^{(--)} \partial_{(--)}) W - dL ,
\end{equation}

 Eq.\p{so(1,1)}
 \begin{equation}\label{3Dso(1,1)}
 d\Om^{(0)} = 1/2 \Om^{--} \Om^{++}
 \end{equation}
 for the forms \p{1NOm--f} -- \p{1NOm0f} results in the equation
\begin{equation}\label{1Nsleq}
\partial_{(--)} D_{(+)} W = - i e^{2W} \Psi^{(+)}_L  ,  \qquad
\end{equation}
 which, together with the identity  \p{ident1} and chirality conditions
 \p{Psich}, describes completely $D=3$ heterotic superstring in the
 framework of geometric approach.

 \bigskip

 Decomposing the superfields in the (finite) power series on
the only nilpotent Grassmann coordinate $\eta^{+}$, we can verify that the
constraint \p{ident1} has no dynamical consequences and simply expresses
the highest components of the superfields
$W$ and $\Psi^{(+)}_L$ through the leading ones
\begin{equation}\label{1Nwexp}
W  = w + {{2i} \over 3} \eta^{(+)} e^{-3w}
\partial_{(++)} (e^{3w}\psi^{(+)}_L),
\qquad
\end{equation}
\begin{equation}\label{1Npsilexp}
\Psi^{(+)}_L  = \psi^{(+)}_L (\xi^{(++)}) + \eta^{(+)}
(1 - 2i \partial_{(++)} \psi^{(+)}_L \psi^{(+)}_L) ,
\end{equation}
The only nontrivial consequence of the
superfield equation \p{1Nsleq} is
\begin{equation}\label{1Nmodleq}
\partial_{(++)}
\partial_{(--)} w = {1 \over 2} e^{2w}
(1 - {{2i} \over 3} \partial_{(++)} \psi^{(+)}_L \psi^{(+)}_L)
\end{equation}
which
can be reduced to the
standard bosonic Liouville equation
\begin{equation}\label{leq}
\partial_{(++)} \partial_{(--)}
\tilde{w} = 1/4 ~ exp\{ 2\tilde{w} \}
\end{equation}
by the field redefinition
\begin{equation}\label{1Nred}
w =
\tilde{w} + {{2i} \over 3} \partial_{++} \psi_L \psi_L
\end{equation}

This, from one hand,  gives us a reason to conclude, that $D=3$ heterotic
string (without heterotic fermions) is described by $n=(1,0)$
supersymmetric extension of the nonlinear Liouville equation and, from the
other hand, means that this nonlinear system is exactly solvable.

Indeed, it can be reduced to the system of the exactly solvable nonlinear
Liouville equation \p{leq} and free field equation
\begin{equation}\label{chirfer}
\partial_{(--)} \psi^{(+)}_L = 0
\end{equation}
for fermionic field.

\bigskip

It is remarkable, that all equations \p{Psich}, \p{ident1}, \p{1Nsleq}
 appears as consequences of zero curvature representation
\begin{equation}\label{MCs1N}
d\Om_{\underline \a}^{~\underline \b} -
\Om_{\underline \a}^{~\underline \g} \Om_{\underline \g}^{~\underline \b}
= 0
\end{equation}
 for $SL(2,R)$ ($=SO(1,2)$) connection
\begin{equation}\label{Oms1N}
\Om_{\underline \a}^{~\underline \b} =  {1 \over 2}
\left(
      \matrix{ \Om^{(0)} & \Om^{--}  \cr
               \Om^{++} & - \Om^{(0)} \cr}
                                          \right)
\end{equation}
 with the forms $\Om^{\pm\pm}, ~\Om^{(0)}$ determined by Eqs.
 \p{1NOm--f}, \p{1NOm++f}, \p{1NOm0f}.

 \bigskip

 In conclusion, let us present
 the B\"{a}cklund transform for $n=(1,0)$ supersymmetric Liouville
system in the superfield form
\begin{eqnarray}\label{1Nback}
D_{(+)} W - D_{(+)} L =
{{-i} \over {a}} exp\{ W+L\} \Psi^{(+)}_L , \qquad \nn
\partial_{(--)} W +
\partial_{(--)} L = a  exp\{ W-L\}  \qquad
\end{eqnarray}
One of the
selfconsistency conditions for \p{1Nback} gives eq.\p{1Nsleq} and another
do not contain $W$ and restrict $L$ by free superfield equation
\begin{equation}\label{1Nfree}
\partial_{--} D_{+} L = 0
\end{equation}

 \begin{center}
 {\bf Conclusion}
 \end{center}

In this talk we have consider in details the previously proposed
\cite{bsv}
generalized action principle on the simple example of $D=10
{}~(and~3, 4, 6)$ heterotic
(super)string,
have proved the off--shell (in rheonomy sense) superdiffeomorphism
invariance of the generalized action,
 and demonstrate that it produce naturally
the torsion constraint of the world--sheet geometry, as well as
the doubly supersymmetric
geometrical approach. The latter had been developed preciously in Ref.
\cite{bpstv} on the basis of the  postulated {\sl a priori}
geometrodynamic equation \p{geomd} and torsion constraints.

But just for the
heterotic string, where the geometrodynamic equation
\p{geomd} do not leads
to any dynamical
equation \cite{dghs92} and, moreover, the superfield form of these
dynamical equations had not bee known, the description
of the minimal embedding
of the world sheet superspace
was unclear \cite{bpstv}.

We have investigated completely the geometric approach equations
for $D=3$, $N=1$ superstring and have proved that they can be reduced to
$n=(1,0)$ supersymmetric generalization
of the nonlinear Liouville equation.

\bigskip

We have stressed that this system of equations appears in the form of zero
curvature representation.

\bigskip

The later property holds for nonlinear equations describing any
super--$p$--brane in the doubly supersymmetric geometric approach.

\bigskip

The consequences of this fact are under investigation now.

\end{document}